\documentclass[amsmat,amssymb,amsfonts,aps,prb,twocolumn,showpacs]{revtex4}

\usepackage{graphicx}
\usepackage{dcolumn}
\usepackage{bm}
\begin{document}

\title{ Ferromagnetism mediated by few electrons in
a semimagnetic quantum dot.}

\author{J. Fern\'andez-Rossier$^1$, L. Brey$^2$}
\affiliation{(1)Departamento de F\'isica Aplicada, Universidad de
Alicante, 03690 San Vicente del Raspeig, Spain \\
(2) Insituto de Ciencia de Materiales de Madrid, CSIC,
Cantoblanco,  28049 Madrid, Spain}

\date{\today}

\begin{abstract}
A  (II,Mn)VI diluted magnetic semiconductor quantum dot with an integer number
of electrons controlled with a gate voltage is considered.  We show that a
single conduction band electron is able to induce a spontaneous collective 
magnetization of the Mn spins, overcoming the short range antiferromagnetic
interactions. The carrier mediated ferromagnetism in the dot survives  at 
temperatures above 1 Kelvin, two orders of magnitude larger than the Curie
temperature for the same material in bulk. The magnetic behavior of the dot
depends dramatically on the parity of the number of injected electrons.

\end{abstract}

 \maketitle


{\em Introduction.}
The range, strength and sign of exchange interactions between magnetic
impurities in diluted magnetic semiconductors (DMS) depends   on the density
and nature of the  states at the Fermi level.  
Mn doped semiconductors of the families (II,Mn)-VI and
(III,Mn)-V order ferromagnetically in the presence of carriers that mediate
indirect exchange interactions between Mn. In the case of (III,Mn)V
compounds like GaAsMn, 
Mn acts as an acceptor supplying holes 
responsible of the ferromagnetism below a transition temperature which depends
on both  Mn and hole densities \cite{Ohno} and can reach 160 K \cite{HTC}. In
contrast, Mn does not supply itinerant carriers in  (II,Mn)VI compounds.  
These materials do not order ferromagnetically \cite{Furdyna}  unless further
doping with acceptors provides holes which produce  ferromagnetism below a
carrier density dependent Curie Temperature ($T_C$) of approximately 2 K
\cite{Haury}.

Electrical  control of the carrier density, in contrast with chemical doping,
has been demonstrated  in a number of DMS heterostructures,   making it
possible to alter reversibly properties of the material like the $T_C$
\cite{Ohno2000,Boukari}    and the coercive field \cite{Ohno2003} of these
systems. The fact that the carrier density is much higher than the relative
change achieved artificially sets limits to the  control.  In contrast, the
number of electrons in a GaAs quantum dot can be varied one by one, starting
from zero,  in  single electron transistors   \cite{set1}. Single electron
transistors with  non magnetic  II-VI (CdSe) quantum dots of lateral size
smaller than 10 nm  have been also fabricated \cite{Klein}. (II,Mn)VI quantum
dots  of similar size have also been  grown and studied magneto-optically by
several groups \cite{DMS-dot}.  Therefore, the fabrication of a single electron
transistor with (II,Mn)VI quantum dots seems feasible and motivates this work.

We study the magnetic properties  of a (II-VI)Mn quantum dot with an integer
number of electrons. The lateral dimensions  of the dot are smaller than 10
nm and, for the range of Mn concentration considered, $x<$0.05, the number of
Mn atoms in the dot is of the order of one hundred and the number of
electrically injected electrons ranges from zero to 10. We find  compelling
theoretical and numerical evidence that the addition of a single electron to an
otherwise paramagnetic DMS dot is enough to couple most of the Mn spins so that
a net total magnetic moment emerges spontaneously. Therefore, a single electron
transistor with a DMS quantum dot would switch on and off the magnetic total
magnetic {\em completely}  in a controlled and reversible manner. 
 
{\em Hamiltonian}.
The Hamiltonian describing the system is the zero dimensional version of the
standard exchange Hamiltonian used   for both  bulk \cite{Furdyna,DMS-bulk} and
two dimensional systems \cite{DMS-2D}. For simplicity we only consider
conduction band electrons, for which we can ignore spin orbit interaction,
and we ignore Coulomb repulsion. The latter   tends to
enhance  carrier mediated exchange interactions so that our results do
not change qualitatively if  this approximation is abandoned. 
Conduction
band electrons (creation operator $c^{\dagger}_{n,\sigma})$  occupy confined
levels $\epsilon^0_n$ of the dot. The Mn spins are described with $S=5/2$
operators $\vec{M}_I$. The spin of the quantum dot electrons  and the Mn
interact via a zero range exchange  interaction. The Mn spins interact also
with each other via a short range antiferromagnetic superexchange
interaction\cite{Furdyna} ${\cal J}_{I,I'}$.
The Hamiltonian reads:
\begin{eqnarray}
{\cal H}= \sum_{n,\sigma}\epsilon^0_{n} c^{\dagger}_{n,\sigma}
c_{n,\sigma}
-
J_c \sum_{I}
\vec{M}_I\cdot \vec{S}_e(\vec{x}_I) + {\cal H}_{AF}
\label{hamil}
\end{eqnarray}
where
$
{\cal H}_{AF}=
\frac{1}{2}\sum_{I,I'}{\cal J}_{II'}\vec{M}_I\cdot\vec{M}_{I'}
$
and 
\begin{equation}
\vec{S}(\vec{x}_I)\equiv \sum_{\sigma,\sigma',n,n'} \phi_n^{*}(\vec{x}_I)
\phi_{n'}(\vec{x}_I)\frac{1}{2}\vec{\tau}_{\sigma,\sigma'}
c^{\dagger}_{n,\sigma}
c_{n',\sigma'}
\end{equation}
is the  quantum dot electron local spin density and $\phi_{n}(\vec{x}_I)$ is
the $n^{th}$ orbital  wave functions of dot. Exchange
interaction produces transitions between different levels of the dot. The
interlevel spacings  considered here are of the order of 20 meV and
higher, except for the degeneracies that some of the dots might have. The 
exchange energy of a quantum dot electron with a single Mn is approximately
given by $j\equiv J_c/\Omega_D$, where $\Omega_D$ is the quantum dot volume.
For Cd$_{1-x}$Mn$_x$Te quantum dots with 10 $nm$ of lateral dimension, 
we have $j\simeq$ 1.5 $\times 10^{-2}$ meV, much smaller than the level spacing.

{\em Effective spin Hamiltonian} .
In the following we derive analitycally the effective interaction between the
Mn spins, which are treated classically. The canonical  ensemble 
equilibrium partition  function  for the dot with a fixed 
number of electrons $N_e$ is given by:  
\begin{equation}
{\cal Z}_N= \int d\vec{M}_1 ..d\vec{M}_N 
e^{-\beta{\cal H}_{AF}}
\sum_{\alpha} e^{-\beta E_{\alpha,N}\left(\vec{M}_I\right)}
\label{part1}
\end{equation}
where $\beta=1/k_BT$ and $E_{\alpha,N_e}\left(\vec{M}_I\right)$ is the energy
of the Slater determinant labeled with $\alpha$,  for a given Mn spin
configuration, $\left(\vec{M}_I\right)$.  At  low temperatures  we can drop
from  the sum over configurations in (\ref{part1}) all except the  ground state
Slater $\alpha=0$, whose energy we denote with 
${\cal E}\equiv E_{0,N_e}\left(\vec{M}_I\right)$. 
In this approximation the effective Mn-Mn coupling is:
\begin{equation}
{\cal H}_{eff}={\cal H}_{AF}+{\cal E}\left(\vec{M}_I\right)
\label{eff}
\end{equation}
Let  $\lambda_l(\vec{M}_I)$ be  the eigenstates of the
Hamiltonian matrix 
\begin{equation}
{\cal H}_{n\sigma,n'\sigma'}\equiv\epsilon^0_{n}
-\frac{J_c}{2}\sum_I \vec{M}_{I}\cdot \vec{\tau}_{\sigma,\sigma'}
\phi_n^{*}(\vec{x}_I)
\phi_{n'}(\vec{x}_I)
\label{matrix}
\end{equation}
associated to  $(\ref{hamil})$ for a given  configuration  of classical
spins, $\vec{M}_I$. This matrix has $n$ intra-level 2$\times$2 boxes. For the
small dots considered here, the leading order term comes from the intra-level
term. To linear order in $J_c$ we have:  
$\lambda_l\simeq\epsilon^0_n \pm 
 \left|\frac{J_c}{2}\sum_{I} |\phi_n(\vec{x}_I)|^2 \vec{M}_I\right|
$. The  many-electron ground state energy  is the sum over the $N_e$ lowest
eigenvalues $\lambda_l$.  
For an odd number of electrons,
$N_e=2N+1$, all the contributions linear in $J_c$ coming from the first $2N$
electrons vanish identically and the only  contribution 
 comes from the most external electron. Modulo an irrelevant constant,
we obtain the following ground state electronic energy for a dot with $2N+1$
electrons:
 \begin{equation}
{\cal E}_{o}=
- \sqrt{\frac{J_c^2}{4} \sum_{I,I'} |\phi_{N+1}(\vec{x}_I)|^2 
|\phi_{N+1}(\vec{x}_{I'})|^2
\vec{M}_I \cdot\vec{M}_{I'} }
\label{Heff1}
\end{equation}
Equations  (\ref{eff}) and (\ref{Heff1}) define the carrier mediated Mn-Mn
coupling which is  one of the important results of the paper.  The Mn-Mn
interaction mediated by an odd number of electrons, including a single
electron, always favors ferromagnetic couplings and it scales  with $J_c$.
These features  are in contrast with the standard bulk  RKKY coupling,
 which scales  with $J_c^2$,  and it can be either positive or
negative. For these  reasons the effective coupling (\ref{Heff1}) is stronger
than its bulk counterpart.   There is yet another difference: RKKY coupling
involves a sum over several orbital states whereas (\ref{Heff1})  only depends
on the most external orbital state, opening the door to a quantum design of
indirect exchange interactions in confined geometries \cite{Mirage}

For an even number of electrons all the contributions linear in $J_c$ cancel.
The leading order contribution to $\cal E_e$ comes from   inter-level exchange
coupling . We  calculate $\cal E_e$ doing perturbation theory around $J_c=0$. 
The magnetic part of the ground state electronic so obtained reads:
\begin{eqnarray}
{\cal E}_{e}
=\sum_{I,I'} \left[\frac{J_c^2}{2}\sum_{n,n'} \gamma_{n,n'}(I,I')
 \frac{f_n-f_{n'}}{\epsilon^0_n-\epsilon^0_{n'}}
\right]\vec{M}_I\cdot\vec{M}_{I'}
\label{Heff2}
\end{eqnarray} 
where $\gamma_{n,n'}(I,I')\equiv \phi^{*}_n(\vec{x}_I) \phi_{n'}(\vec{x}_{I})
\phi^{*}_n(\vec{x}_{I'}) \phi_{n'}(\vec{x}_{I}')$ and $f_n=0,1$  are the
occupation of  the unperturbed dot in the ground state electronic
configuration.   The effective couplings (\ref{Heff2}) are weaker than 
(\ref{Heff1}) and can be both  positive and negative (ferro or
antiferromagnetic) for a given dot and different Mn couples. The striking
differences between effective interactions (\ref{Heff1}) and (\ref{Heff2})
permit to anticipate very different behavior  for dots with open and closed
shell electronic structure.

{\em Local mean field theory}.
Effective interactions (\ref{Heff1}) and (\ref{Heff2}) between the spins result
from integrating out the conduction electrons in some limits and treating the
Mn spins classically. We now do a mean field theory for Hamiltonian
(\ref{hamil}), keeping track of both Mn and electrons. Quantum dot electrons
interact with an effective exchange field provided by the expectation value of
the Mn spin operators. The latter is calculated assuming that each Mn spin
interact with an effective field provided by the quantum dot spin density {\em
and} its neighboring   Mn spins, via antiferromagnetic superexchange
interaction. Since both the electron spin density and the Mn neighborhood are
different for each Mn, each Mn has a different expectation value, in contrast
with the homogeneous models used for    bulk  \cite{Furdyna,DMS-bulk}. The
local mean field felt by the Mn at $\vec{x}_I$ reads:
\begin{equation}
g\mu_B\langle{\cal \vec{B}}_I\rangle=J_c \langle \vec{S}(\vec{x}_I)\rangle-
\frac{1}{2}\sum_{I'}{\cal J}_{II'}\langle \vec{M}_{I'} \rangle
\label{SC1}
\end{equation}
The expectation value for the magnetization of a spin $S=5/2$ in the effective
field of eq. (\ref{SC1}) reads:
\begin{equation}
\langle \vec{M}_I \rangle = S \vec{u}_I B_S\left(\frac{S}{k_B T}
 g\mu_B\langle{\cal \vec{B}}_I\rangle\right) 
\label{SC2}
\end{equation}
where $\vec{u}_I$ the unitary vector parallel to $\langle{\cal
\vec{B}}_I\rangle$ and $B_S$ is the Brillouin function \cite{Furdyna}.
Finally, the equation for the average conduction electron local
spin density is:
\begin{equation}
 \langle \vec{S}(\vec{x}_I)\rangle=
\frac{  1}{Z_{el}}
 \sum_{\alpha} 
 e^{-\beta E_{\alpha}}
\sum_{l}\langle \Phi_l|\vec{S}(\vec{x}_I) |\Phi_l\rangle 
f_l(\alpha)
\label{SC3}
\end{equation}
where $Z_{el}=\sum_{\alpha}e^{-\beta E_{\alpha}}$ and $|\Phi_l\rangle$ is
eigenvector associated to the $\lambda_l$  eigenvalue of Hamiltonian
\ref{matrix} with $\vec{M}_I$ replaced by $\langle \vec{M}_I \rangle$ given by
eq.  (\ref{SC2}) and $f_l(\alpha)$ is the occupation (0 or 1)  level  in the
many electron configuration $\alpha$.

In the following we show results for  Cd$_{1-x }$Mn$_x$Te hard wall cubic
quantum dots of dimensions $L_x$, $L_y$ and $L_z$ for which $\epsilon^0_n$ and
$\phi_n$ are known analytically. Following reference \onlinecite{Furdyna} we
take $J_c=15$ $eV\AA^3$ . Superexchange antiferromagnetic coupling ${\cal
J}_{II'}$ decays  exponentially with distance and we  only consider coupling
to  first neighbours,  ${\cal J }_{FN}=$ 0.5 meV \cite{Furdyna,AF}.  
The positions of the Mn atoms are randomly chosen  in the cation sites of the
Zinc Blende  lattice with lattice constant, $a=6.4 \AA$, avoiding double
occupation of the sites. The initial value of $\langle \vec{M}_I\rangle$ is
also chosen randomly. Self consistent numerical solution of  equations
(\ref{SC1}),(\ref{SC2}) and (\ref{SC3}) yield converged values of 
$\langle \vec{M}_I\rangle$
which are stable, independent of all  initial conditions, convergence procedure
and small variations of the energy scales of the problem. We characterize  the
collective magnetic order with the average Mn magnetization per Mn:
\begin{equation}
\langle M \rangle \equiv 
\sqrt{ \left|\frac{1}{N}\sum_{I} \langle
\vec{M}_I \rangle\right|^2}  
\label{OP1}
\end{equation}
which can range between $0$ and $S=5/2$. 
Different realizations of the Mn positional  configurations, $\{\vec{r}_I\}$
for otherwise identical dots give different values of $\langle M\rangle$. 
Experiments with a single dot are feasible in principle, so that $\langle
M\rangle$ is observable for a single dot. To make sure that general results are
independent of a particular Mn realization we perform averages over different
realizations of $\{\vec{r}_I\}$. The net magnetization of a dot  averaged  over
configurations is denoted by $\langle \langle M \rangle \rangle$.  The mean 
standard deviation among different realizations of $\{\vec{r}_I\}$ is denoted
by $\sigma_M$.
 \begin{figure}
[hbt]
\includegraphics[width=3.375in]{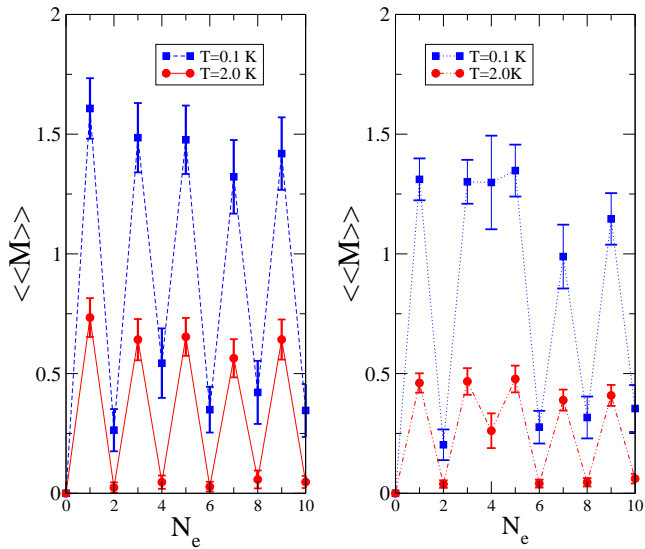}
\caption{ \label{fig1}(Color online). Magnetization per Mn versus $N_e$, averaged over disorder
configuration for two different dots. }
\end{figure}

 In figure 1 we plot $\langle \langle M \rangle \rangle$ and $\sigma_M$
(vertical bars) as a function of $N_e$ for two different dots
 at two different temperatures, $k_BT$=0.1 K and 2.0 K.  Dots 1 and 2 (left and
right panels) have the same Mn concentration  and the same
 $L_x$  ($x=0.01$, $L_x$=4 nm) but
different dimensions. Dot 1 has $L_y=$ 6 nm,$L_z$=7 nm and dot 2 has $L_y=$8.5
nm,$L_z=$ 9nm. The number of Mn impurities is 25 and 46 respectively.   We find
that: 

{\em i)} A single electron is enough to induce a spontaneous collective
magnetization different from zero in the absence of applied field. The
collective magnetization survives at temperatures of 1 Kelvin and higher. The
bulk  mean field Curie Temperature  for parabolic bands yields, for
Cd$_{0.99}$Mn$_{0.01}$Te,
$k_B T_C|_{\rm Bulk} = \frac{S(S+1)}{3}\frac{3}{2}c_{\rm Mn}
\frac{n}{\epsilon_F}\simeq 20$ mK 
 where $n$ and $c_{\rm Mn}$ are
the electron and Mn density respectively corresponding to the dot of figure 1.
Therefore,  carrier mediated spin correlations in the 
 quantum dots survive at
temperatures 2 orders or magnitude larger than  $T_C$ in bulk.

{\em ii)}  The addition or removal of a single electron  produces a dramatic
change in the magnetization of the dots.

{\em iii)} The spontaneous magnetization is larger, in general, 
for open shell configurations (odd number of electrons) than
for closed shell configurations  (even number of electrons). Closed shell
configurations with a small gap can yield larger inter-level couplings. 
This is the case of $N_e=4$, specially in  dot 2  with smaller
energy level spacings than  dot 1.

\begin{figure}
[hbt]
\includegraphics[width=3.375in]{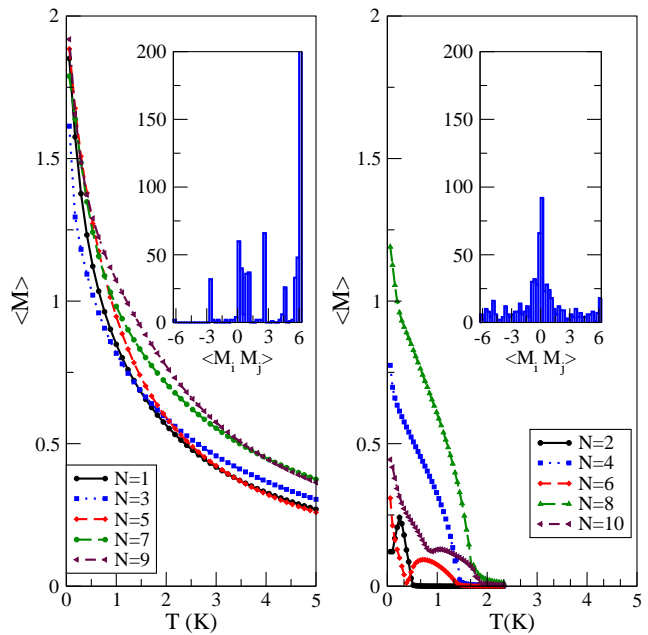}
\caption{ \label{fig2}  (Color online). Magnetization per Mn vs temperature for odd (left) and
even (right) number
of electrons for dot 1 (see text).  In the inset we show
the histograms of $\langle \vec{M}_i\cdot \vec{M}_j
\rangle$ at $k_b T=50$mK for N=1 (left panel)   and
N=2 right panel}
\end{figure}

In figure 2 we show $M(T)$ for a single Mn realization for  dot 1. 
In  the left (right) panel of fig. 2 we show $M(T)$  for an odd
(even) number of electrons. The shape of the curves for $N_e$ odd
 are all
very similar to each other and resemble the spontaneous magnetization of a
confined exciton-polaron \cite{Bhatta}.   In contrast, the shape for the 
$M(T)$ curves varies from case to case  for  even
$N_e$ and  results from the complicated competition of
carrier mediated ferro and antiferromagnetic couplings and superexchange.  To 
quantify this, we plot in the insets  histograms of the values of the
correlation matrix $\langle \vec{M}_I \cdot\vec{M_J}\rangle$ for 
$N_e=1$ (left) and $N_e=2$ (right) at very low temperatures (50 mK). The 
$N_e=1$ histogram is biased towards positive values (ferromagnetic couplings)
whereas  the $N_e=2$  histogram  displays a rather symmetric distribution of
ferromagnetic and antiferromagnetic couplings.

\begin{figure}
\includegraphics[width=2.4in]{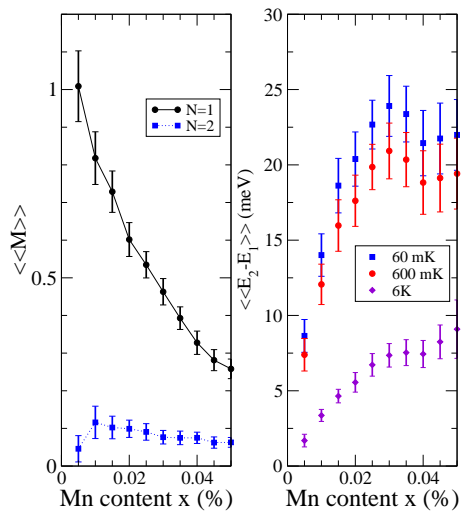}
\caption{ \label{fig3} (Color online). Left panel: $\langle\langle M\rangle \rangle$ per Mn
 as a function of Mn content, $x$ at $k_BT =0.6$ Kelvin. Right panel: spontaneous
spin splitting of the lowest level of the dot, at three temperatures, as a
function of Mn content $x$.}
\end{figure}

The competition between carrier mediated and superexchange couplings evolves as
the Mn concentration increases. In the left panel of figure 3 we show
$\langle\langle M\rangle \rangle$ and $\sigma_M$ (vertical bars) as a function
of Mn concentration, $x$, for dot 2.  We see that, for $N_e=1$,  
$\langle\langle M\rangle \rangle$ decreases monotonically as a function of $x$
as a result of the increase of the number first neighbour pairs coupled
antiferromagnetically. The curve for $N_e=2$  has a maximum around $x=0.01$ and
decreases at higher concentrations. 

The spin rotational symmetry of Hamiltonian (\ref{hamil}) makes any direction
in the spin space equivalent. Therefore, when a spontaneous magnetic moment
arises from  carrier mediated Mn-Mn interactions, it can point in any direction
of the space. The effects related to spontaneous magnetization can be observed
provided that  the collective magnetization  is static during the probe time
scale. A small source of spin anisotropy, like an applied field or Rashba spin
orbit interaction,  can be very efficient in slowing down the magnetization
dynamics.  The spontaneous magnetization results in a splitting of the quantum
dot energy levels which could be measured in transport\cite{set1}.  In the
right panel of fig. 3 we plot the energy splitting of the lowest quantum dot
doublet  as a function of $x$, for $N_e=1$ at three different temperatures for
dot 2. The splitting is a
decreasing function of temperature and 
an increasing function of the Mn content up to
$x=0.03$,  declining for higher $x$, due to the increase 
of first neighbours pairs.

In summary, our results indicate that a very large control of  carrier mediated
interactions can be achieved in quantum dots in which the number of electrons
can be changed one by one.   We claim that a few conduction band electrons
couple the spin of several tens of Mn atoms in a (II,Mn)VI semiconductor
quantum dot.  An odd number of electrons, including just one, yield a
ferromagnetic coupling (eq. \ref{Heff1}) and left panels of figures 1 and 2)
whereas and even number of electrons give both ferro and antiferromagnetic
carrier mediated couplings (eq. \ref{Heff2} and right panels of figures 1 and
2). The result of the competition between carrier mediated interactions and
short range antiferromagnetic superexchange is  a spontaneous collective
magnetization which survives at temperatures of the order of 1 Kelvin, 2 orders
of magnitude higher than the mean field prediction for the same  material in
bulk.
The resulting spontaneous  spin splitting of
the energy levels of the dot  could be measured in transport experiments.

We acknowledge fruitful discussions with J.J. Palacios and E. Louis. Financial
support is acknowledged from Grants No MAT2002-04429-c03-01,
MAT2003-08109-C02-01, Ramon y Cajal Program (MCyT,Spain), Fundaci\'on Ram\'on
Areces,  and   UA/GRE03-15.  This work  has been
partly funded by FEDER funds.


\widetext
\end{document}